\documentclass[epj,final]{svjour}

\usepackage{latexsym}
\usepackage{url}
\usepackage{amsfonts}
\usepackage{amsmath, amssymb}
\RequirePackage{graphicx}

\begin{document}
\title{Black hole shadow to probe modified gravity}
\author{A.~Stepanian\inst{1}, Sh.~Khlghatyan\inst{1}, V.G. Gurzadyan\inst{1,2}
}                     
%
%
\institute{Center for Cosmology and Astrophysics, Alikhanian National Laboratory and Yerevan State University, Yerevan, Armenia \and
SIA, Sapienza Universita di Roma, Rome, Italy}
\date{Received: date / Revised version: date}
%

\abstract{We study the black hole's shadow for Schwarzschild-de Sitter and Kerr-de Sitter metrics with the contribution of the cosmological constant $\Lambda$. Based on the reported parameters of the M87* black hole shadow we obtain constraints for the $\Lambda$ and show the agreement with the cosmological data. It is shown that, the coupling of the $\Lambda$-term with the spin parameter reveals peculiarities for the photon spheres and hence for the shadows. Within the parametrized post-Newtonian formalism the constraint for the corresponding $\Lambda$-determined parameter is obtained.}

\PACS{
      {98.80.-k}{Cosmology}   
     } 
%
\maketitle

\section{Introduction}

The release of the M87* massive black hole (BH) shadow image by the EHT Collaboration \cite{M87a,M87a1,M87a2,M87b} marked a new phase of study of a number of physical effects occurring in the very centers of galactic nuclei. The BH's shadow image is used to constraint both the BH parameters including its spin within General Relativity (GR), as well as the properties of the accretion disk, see \cite{M87c,M87d,M87e} and references therein. The shadow parameters appear to be efficient also in testing for strong-field conditions the parametrized post-Newtonian (PPN) formalism in first order approximation \cite{PPN}.
  
In this Letter we consider the possibility to use the M87* shadow available information to constraint the Schwarzschild-de Sitter (SdS) and Kerr-de Sitter (KdS) BHs, i.e. taking into account the cosmological constant $\Lambda$ in the metric. The $\Lambda$-term in the spherically symmetric metric is arising also in the modified weak-field General Relativity (GR) which enables to consider the common nature of the dark energy and dark matter \cite{G,GS1,GS2,GS3}.   
That approach also suggests a test for the potential deviations from GR having in view of the value of the cosmological constant $\Lambda = 1.11 \times 10^{-52} m^{-2}$ \cite{Pl} using the gravity lensing observations \cite{GSlens}.

We use key properties of the photon orbits in the presence of non-zero $\Lambda$ for SdS and KdS metrics and of the shadows, to obtain the constraints on the numerical value of $\Lambda$.  We also analyze the role of the $\Lambda$ term for the shadow properties within the parametrized post-Newtonian formalism. 

\section{Constraining the value of $\Lambda$ from the shadow}

The Schwarzschild-de Sitter metric, i.e. the spherically symmetric metric with non-vanishing $\Lambda$-term has the following form \cite{R}
\begin{equation}\label{metric}
ds^2=\left( 1-\frac{2 GM}{r c^2} - \frac{\Lambda r^2}{3} \right) c^2 dt^2- \left( 1-\frac{2 GM}{r c^2} - \frac{\Lambda r^2}{3} \right)^{-1} dr^2 - r^2 d\Omega^2.
\end{equation}
Importantly, it has been shown that the $\Lambda$-term does not affect the null geodesics \cite{Is,Con1,L}. Thus, it can be checked that, comparing to $\Lambda =0$ case the photon sphere will remain unchanged i.e. 
\begin{equation}\label{psphs}
r_{ph} = \sqrt{g_{tt}}\left(\frac{d \sqrt{g_{tt}}}{dr}|_{r_{ph}}\right)^{-1} = \frac{3GM}{c^2}.
\end{equation}
However, due to the presence of the $\Lambda$ term in the $g_{tt}$ component, the same is not true for the radius of the BH's shadow which is equal to
\begin{equation}
r_{sh} = \frac{r_{ph}}{\sqrt{g_{tt}(r_{ph})}}.
\end{equation}
In this sense, comparing to Schwarzschild metric where the shadow is 
\begin{equation}
r_{sh,sh} = 3\sqrt{3}\frac{GM}{c^2}, 
\end{equation}
we get
\begin{equation}\label{ShL}
r_{sh,\Lambda} = \frac{\frac{3GM}{c^2}}{\sqrt{\frac{1}{3} - 3 (\frac{GM}{c^2})^2 \Lambda}}.
\end{equation}
Thus, by taking the recently reported values of M87* BH shadow we can find constraints on the numerical value of the cosmological constant. Namely, the upper limits for $\Lambda$ can be obtained based on two different reported values. 

First, we get the upper limit based on the numerical values for  $r_{sh} = 42 \pm 3 \mu as$. In this case the constraint is equal to
\begin{equation}
1 + \frac{\mathbb E(r_{sh})}{r_{sh}} \geq \frac{r_{sh,\Lambda}}{r_{sh,sh}} = \frac{1}{\sqrt{1-9\left(\frac{GM}{c^2}\right)^2 \Lambda}}.
\end{equation}
Next, we repeat the same analysis this time by considering the mass of M87* i.e. $M = (6.5 \pm 0.7) \times 10^9 M_{\odot}$.
It should be noticed that we cannot take into account both errors simultaneously, since the mass of the BH and the radius of the shadow are dependent on each other. Consequently, for these two cases we get
\begin{equation}
\Lambda \leq 1.542 \times 10^{-28} m^{-2}, \quad \Lambda \leq 2.214 \times 10^{-28} m^{-2}.
\end{equation}
The obtained limits, as we see, are close to each other and are in agreement with the cosmological value of $\Lambda$. 

\section{Schwarzschild-de Sitter vs Kerr-de Sitter BHs}

For Schwarzschild-de Sitter metric in Eq.(\ref{ShL}) we get $r_{sh,\Lambda} \to \infty$ once 
\begin{equation}\label{con}
9 \left(\frac{GM}{c^2}\right)^2 \Lambda =1.
\end{equation}
It can be checked that, this relation is also the condition of having the so-called extreme SdS BH solution. Indeed, it is known the event horizons of the SdS metric i.e. Eq.(\ref{metric}) are equal to
\begin{equation}\label{EH}
r_1 = \frac{2}{\sqrt \Lambda} cos\left(\frac{1}{3} cos^{-1}\left(\frac{3GM \sqrt \Lambda}{c^2}\right)+\frac{\pi}{3}\right), \quad r_2 = \frac{2}{\sqrt \Lambda} cos\left(\frac{1}{3} cos^{-1}\left(\frac{3GM \sqrt \Lambda}{c^2}\right)-\frac{\pi}{3}\right), \quad r_3 = -(r_1 + r_2).
\end{equation}
However, once the condition in Eq.(\ref{con}) is satisfied,  instead of two (positive and real) event horizons we will have one horizon of radius equal to
\begin{equation}
r_1 = r_2 = \frac{1}{\sqrt{\Lambda}}.
\end{equation}
Interestingly, one can check that at $r=\frac{1}{\sqrt{\Lambda}}$ the gravitational attraction of Newtonian term in Eq.(\ref{metric}) will be completely balanced by the repulsive force produced by $\Lambda$ term i.e.
\begin{equation}\label{bal}
\left(\frac{GM}{r^2} - \frac{\Lambda c^2 r}{3}\right) \bigg|_{r=\frac{1}{\sqrt{\Lambda}}} = 0.
\end{equation}
Moreover, it is commonly believed that in more realistic astrophysical cases a BH is described not by Schwarzschild metric, but by Kerr metric, where the spin parameter of BH i.e. $a = \frac{J}{Mc}$ is also taken into account. Indeed, the presence of $a$ as the indicator of BH's intrinsic angular momentum leads to several interesting results which have been studied extensively in the literature (see e.g. \cite{K1,K2,K3}). Accordingly, for Kerr BH instead of one photon orbit, one will have two of them with the following radii
\begin{equation}\label{psphK}
r_1 = \frac{2 GM}{c^2} (1 + cos(\frac{2}{3} cos^{-1} (- \frac{|a|c^2}{GM}))), \quad r_2 = \frac{2 GM}{c^2} (1 + cos(\frac{2}{3} cos^{-1} (+ \frac{|a|c^2}{GM}))).
\end{equation} 
Clearly, for $a=0$ the two solutions will coincide and the solution of Schwarzschild case in Eq (\ref{psphs}) is recovered. Considering our analysis, here we are interested to include the $\Lambda$ in the metric of rotating BHs. Namely, in our case this will be the Kerr-de Sitter (KdS) metric which is defined as follows
\begin{equation} \label{KdS}
ds^2 = \frac{\Delta_r}{\rho^2 L^2}\left(cdt-a\sin^2\theta d\phi\right)^{2} - \frac{\rho^2}{\Delta_r} dr^2 - \frac{\rho^2}{\Delta_\theta} d\theta^2 - \frac{\Delta_\theta \sin ^2 \theta}{\rho^2 L ^2} \left(a cdt - (r^2+a^2)d\phi\right)^{2}, 
\end{equation}
where
\begin{equation} 
\begin{aligned}
	&\Delta_r = \left(1- \frac{\Lambda r^2}{3}\right)(r^2+a^2) - \frac{2 G M r}{c^2}, \\
	&\Delta_\theta = \left(1+ \frac{a^2 \Lambda \cos^2 \theta}{3}\right), \\
	&L = \left(1+ \frac{a^2 \Lambda}{3}\right), \\
	&\rho^2 = r^2+ a^2 \cos ^2 \theta.
\end{aligned}
\end{equation}
In this sense, KdS metric describes the geometry of spacetime when a single axially symmetric object is immersed in de Sitter background \cite{KdS1,KdS2,KdS3}. As a result, the photon sphere for KdS metric can be obtained by solving the following cubic equation
\begin{equation}\label{psph}
3\left(1+\frac{1}{3}\Lambda a^2\right)^2r^3 + 6(\frac{GM}{c^2})\left(\Lambda a^2 -3\right)r^2 + 27 (\frac{GM}{c^2})^2r - 12\left(\frac{GM}{c^2}\right)a^2 = 0 
\end{equation}
The key point of the above equation is the coupling of $a$ and $\Lambda$. Namely, we have no free $\Lambda$-term which means that for $a=0$ the equation will be reduced to standard Schwarzschild case. Clearly, this fact illustrates that for spherically symmetric BHs, no matter $\Lambda$ is vanishing or not, the photon sphere will be equal to $3\frac{GM}{c^2}$. But in axially symmetric case we have the coupling of $\Lambda$ term with spin parameter and as a result of that in contrast to spherically symmetric case, the photon sphere of Kerr and KdS BHs will be different.  

However, the differences for Kerr and KdS BHs for astrophysical configurations such as M87* are too small to be detected via current observational methods. In fact, the difference between Eq.(\ref{psph}) and pure Kerr case is the presence of $\frac{1}{3}\Lambda a^2$ and $\Lambda a^2$ in the 3rd and 2nd degrees of the polynomial, respectively. Similarly, the difference between Kerr and Schwarzschild case is arisen due to $12\left(\frac{GM}{c^2}\right)a^2$. Nevertheless, the main point is that, while considering the current value of cosmological constant and the parameters of a BH such as M87*, it turns out the contribution of $\Lambda a^2$ is too small to be observed at the typical astrophysical scales. In particular, for M87* this contribution is around $10^{-27}$ which is far smaller than $\frac{GM}{c^2}a^2$ in the last term of Eq.(\ref{psph}). 

Here, the essential point is that, the nature of photon orbits in both Kerr and KdS BHs is identical. Indeed, since the roots of Eq.(\ref{psph}) can be considered as the radii of photon orbits, finding the real and positive roots is of the main importance. Meantime, it should be noticed that the number of real and positive solutions depends on the sign of $\Lambda a^2 -3$. Based on the current astrophysical data for a BH of M87* type this value will be definitely negative. Thus, one can state that there are three positive solutions. Furthermore, it can be shown that the value of the first positive root is smaller than the radius of the outer event horizon of BH. Namely, for the data of M87* the solutions of full Eq.(\ref{psph}) for radii in the presence of positive $\Lambda$, are (in meters)
\begin{equation}\label{Solph}
r_1 = 5.124 \times 10^{12}, \quad r_2 = 1.5 \times 10^{13}  ,\quad r_3= 3.767 \times 10^{13}.
\end{equation}
On the other hand, by taking the metric of KdS according to Eq.(\ref{KdS}) it becomes clear that the event horizons of BH are obtained by solving the following equation
\begin{equation}\label{KdSEH}
\Delta_r = \left(1- \frac{\Lambda r^2}{3}\right)(r^2+a^2) - \frac{2 G M r}{c^2} = 0.
\end{equation}
Consequently, for M87* the radii of event horizons (in meters) will be 
\begin{equation}\label{EHNum}
EH_1 = -1.643 \times 10^{26}, \quad EH_2 = 5.434 \times 10^{12} , \quad EH_3 = 1.383 \times 10^{13}, \quad EH_4 = 1.643 \times 10^{26}.
\end{equation}
Following the geometry of KdS BH it turns out that the $EH_4$ is the so-called ``cosmological horizon" and is located far beyond the other two horizons i.e. $EH_2$ and $EH_3$ which are regarded as the inner and outer event horizons of KdS in the same analogy with Kerr BH. Meantime, the negative $EH_1$ is interpreted as the dual of $EH_4$ which is located on the other side of BH's ring singularity and becomes important only during the KdS solution's mathematical extension.

Accordingly, by comparing the Eqs.(\ref{Solph},\ref{EHNum}) we find
\begin{equation}\label{orbits}
r_1 < EH_3 < r_2, r_3 < EH_4.
\end{equation}

Thus, similar to the Kerr case, here two photon orbits will be formed outside the BH. Indeed, the presence of these two solutions for Kerr BH can be regarded as a manifestation of frame dragging effect. Namely, the frame dragging for KdS metric in Eq.(\ref{KdS}) is defined as 
\begin{equation}\label{fd}
\Omega = - \frac{g_{t\phi}}{g_{\phi \phi}},
\end{equation}
so while the photon at inner orbit i.e. $r_{2}$ moves in the same direction of BH's spin, the motion at the outer orbit $r_3$ is in the opposite direction which is due to the presence of BH's intrinsic spin. In this sense, we can conclude that the presence of $\Lambda$ term in the BH equations only corrects the radii of photon orbits  while the nature of the frame dragging effect itself and particularly the formation of photon orbits remains intact. Finally, since the well-known Lense-Thirring (LT) precession can be obtained from Eq.(\ref{fd}) in the slow rotation limit, our above statement about the effect of $\Lambda$ on the frame dragging effect can be regarded as the continuation of investigation of LT precession in the presence of $\Lambda$ \cite{SKG}. It was shown that then an additional term is appeared which contains $\Lambda$ 
\begin{equation}\label{LTL}
\Omega_{LT} = \frac{2 GJ}{c^2 r^3} + \frac{\Lambda J}{3 M},
\end{equation}
and which can be interpreted as a correction to the so-called gravito–gyromagnetic ratio.
However, in the above relation the $\Lambda$ is coupled to the mass and angular momentum of the rotating object. In other words, in the presence of a positive $\Lambda$ the LT precession is corrected according to Eq.(\ref{LTL}) and no new effect is reported. This is due to the fact that, without the rotation of the central object the $\Lambda$ itself cannot cause any precession at all.

In \cite{PPN} it has been proposed that based on the data of BH's shadow one can get constraints on the variables of parametrized post-Newtonian formalism. Namely, by considering the expansion to the $r^{-3}$ order term one has
\begin{equation}
g_{tt} = 1- \frac{2GM}{c^2 r} + \frac{2 (\beta - \gamma) (GM)^2}{c^4 r^2} - \frac{2 \zeta (GM)^3}{c^6 r^3}.
\end{equation}
Accordingly, since $\beta - \gamma \approx 0$, the $r_{ph}$ and $r_{sh}$ in the $M=c=G=1$ units will be written as
\begin{equation}\label{shzeta}
r_{ph} = 3 + \frac{5}{9}\hat \zeta, \quad r_{sh} = 3\sqrt3\left(1+\frac{1}{9}\hat \zeta\right).
\end{equation}
By substituting the numerical values, we find $$\hat\zeta_{\Lambda} = 4.77 \times 10^{-26},$$
as the order of the potential discrepancy from the standard GR.
The value of $\hat\zeta$ is a measure of deviations from GR and it is also a criterion to put constraints on the free parameters of different modified theories of gravity. 

\section{Conclusions}

We studied the effect of the cosmological constant $\Lambda$ on the shadow of BH for Schwarzschild-de Sitter and Kerr-de Sitter metrics. The recently reported data of M87* BH shadow was used to obtain the constraints on the numerical value of $\Lambda$. Namely, two upper limits have been obtained which are based on the error limits of the shadow and the BH mass. Comparing both limits with the current value of $\Lambda$ we have shown that there is no inconsistency among those. It should be noticed that, the importance of such kind of analysis is not to directly observe the effect of the modified term, but to check the validity of the modified theory of gravity according to reported data. 

Then, it is revealed that the condition for having the extreme case for BH shadow is equivalent to the formation of extreme SdS BH. Furthermore, we analyzed the structure of the Kerr BH in the presence of non-zero $\Lambda$. It is shown that while in case of spherically symmetric BHs the radius of photon sphere remains unchanged for both Schwarzschild and SdS BHs, for axially symmetric case due to the coupling of spin parameter $a$ with $\Lambda$ the radii are changed. However, similar to Lense-Thirring precession, the nature of the frame dragging effect does not change in the presence of $\Lambda$. Namely, the radii are modified and again as in the Kerr BH case, two photon orbits i.e. one prograde and other retrograde, are formed.

We also checked the potential numerical deviation due to the $\Lambda$-term from standard GR based on the PPN formalism. We obtained the corresponding $\hat\zeta_{\Lambda}$ parameter, which although being too small to be observed, can be regarded as an indication of deviation from GR.

\end{document}